\documentclass[conference]{IEEEtran}
\IEEEoverridecommandlockouts
\usepackage{cite}
\usepackage{amsmath,amssymb,amsfonts}
\usepackage{algorithmic}
\usepackage{graphicx}
\usepackage{textcomp}
\usepackage{xcolor}
\usepackage{mathtools}
\usepackage[pscoord]{eso-pic}
\def\BibTeX{{\rm B\kern-.05em{\sc i\kern-.025em b}\kern-.08em
    T\kern-.1667em\lower.7ex\hbox{E}\kern-.125emX}}

\newcommand\blfootnote[1]{%
  \begingroup
  \renewcommand\thefootnote{}\footnote{#1}%
  \addtocounter{footnote}{-1}%
  \endgroup
}
\begin{document}
\title{AutoCkt: Deep Reinforcement Learning of Analog Circuit Designs\\
}

\author{\IEEEauthorblockN{Keertana Settaluri, Ameer Haj-Ali, Qijing Huang, Kourosh Hakhamaneshi, Borivoje Nikolic}
\IEEEauthorblockA{University of California, Berkeley}
\IEEEauthorblockA{\{ksettaluri6,ameerh,qijing.huang,kourosh\_hakhamaneshi,bora\}@berkeley.edu}
}
\maketitle

\begin{abstract}
Domain specialization under energy constraints in deeply-scaled CMOS has been driving the need for agile development of Systems on a Chip (SoCs). While digital subsystems have design flows that are conducive to rapid iterations from specification to layout, analog and mixed-signal modules face the challenge of a long human-in-the-middle iteration loop that requires expert intuition to verify that post-layout circuit parameters meet the original design specification. Existing automated solutions that optimize circuit parameters for a given target design specification have limitations of being schematic-only, inaccurate, sample-inefficient or not generalizable. This work presents AutoCkt, a machine learning optimization framework trained using deep reinforcement learning that not only finds post-layout circuit parameters for a given target specification, but also gains knowledge about the entire design space through a sparse subsampling technique. Our results show that for multiple circuit topologies, AutoCkt is able to converge and meet all target specifications on at least 96.3\% of tested design goals in schematic simulation, on average 40$\times$ faster than a traditional genetic algorithm. Using the Berkeley Analog Generator, AutoCkt is able to design 40 LVS passed operational amplifiers in 68 hours, 9.6$\times$ faster than the state-of-the-art when considering layout parasitics.
\end{abstract}


\section{Introduction}
As technology nodes scale, it becomes increasingly difficult to bring innovation to circuit systems. Because of the complexity of design rules and prominence of layout parasitics in advanced processes, significant design time has to be allocated in order for modern circuits to be taped out. Traditionally, this design time falls to human circuit designers, who are heavily involved in the process of creating these circuit systems. The process of finding circuit parameters to meet a given target design specification heavily relies upon the expert circuit designer to create equations and iterate through values until converging to a solution. In order to reduce time-to-market, it therefore becomes crucial to identify and automate time consuming procedures in a simulation efficient and accurate manner. 

Prior techniques for automating circuit synthesis can be categorized into knowledge-based and optimization-based approaches\cite{Barros2010}. Knowledge-based approaches consist of transcribing circuit knowledge into programs \cite{Jangkrajarng2003,Zhang2006}. These algorithms encapsulate the designer's knowledge through equations, but a large overhead is required for defining any new design, including the time consuming process of hand-crafting equations.

\begin{figure}[!t]
\centerline{\includegraphics[width=3.33 in]{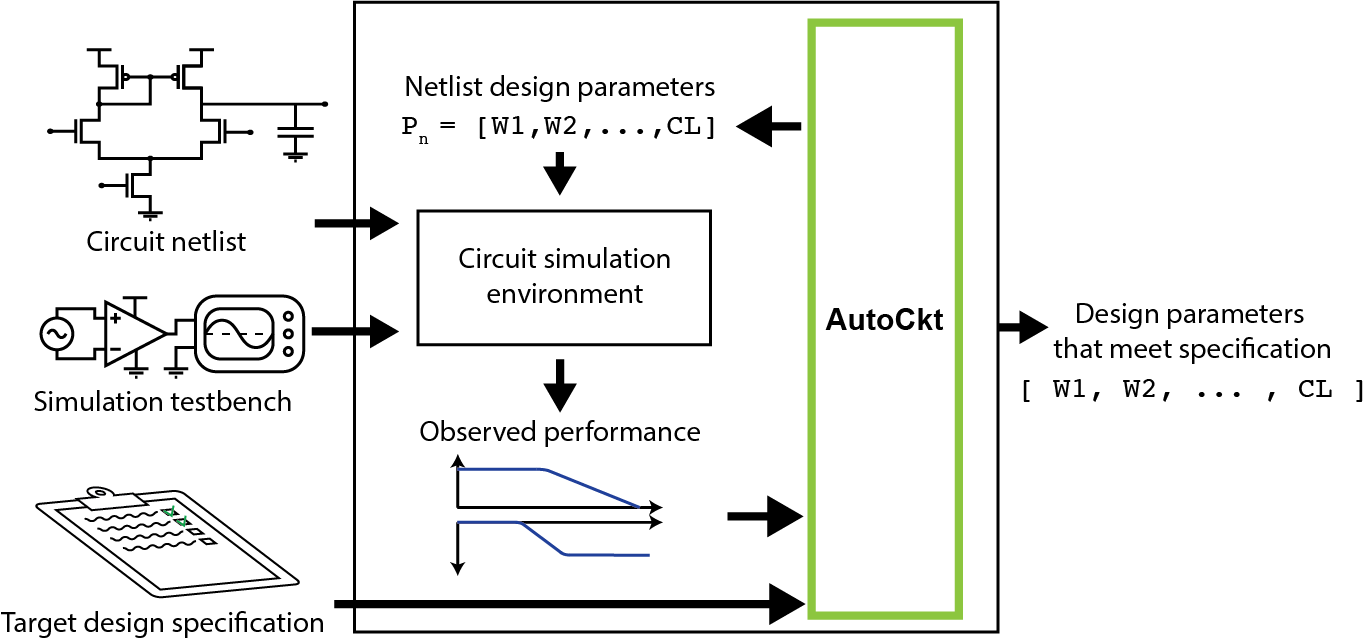}}
\caption{Top level overview, showing what information is needed for AutoCkt in order to design any circuit topology to meet a given target design specification}
\label{system}
\end{figure}
Optimization-based approaches are split into three main sub-categories: equation-based, simulation-based, and learning-based methods. Equation-based methods like geometric programming \cite{Daems2002} manually or automatically obtain constraint equations to then solve and optimize. Though the solvers are quite efficient, creating equations takes time, and only a few predefined circuits can be characterized in this way.

Simulation-based approaches such as genetic algorithms have been explored in depth \cite{Cohen2015}. They function by stochastically sampling an initial population and mutating the best children to produce offspring to then simulate and sample from again. Traditionally, these methods are sample inefficient, and not guaranteed to converge because of stochasticity. In addition, they require re-starting the algorithm from scratch if any change is made to the goal.

Learning-based tools use machine learning methods to solve the analog design problem. In particular, prior work focuses on the usage of supervised or reinforcement learning to determine the relationship between design specification and parameter output. \cite{Wang} uses reinforcement learning to create an agent that traverses through the design space to converge to parameters that meet a particular design specification. The algorithm, however, must be re-trained from scratch every time a design specification changes, which makes this approach extremely sample inefficient. Furthermore, they do not consider layout parasitics. \cite{Hakhamaneshi2019} accelerates the genetic algorithm optimization process by having a deep neural network discriminate against weaker generated samples. In this space \cite{Hakhamaneshi2019} appears to be the most sample efficient algorithm to date.

Other tools that size circuits while considering layout parasitics also exist \cite{Habal2011,Castro-Lopez2008}. Despite improving accuracy compared to schematic-only simulations, they are either inaccurate because they use an approximate parasitics model to speed up simulation time, or use a lookup table a-priori to simulate all relevant designs, making them sample inefficient and time consuming.

In summary, there is a need for a sample efficient, accurate, generalizable and intuitive method for solving analog circuit sizing without the overhead of constraint generation. 

\subsection{Our Contributions}

Inspired by the sequential thought process used by expert analog designers, we present AutoCkt, a machine learning framework to solve analog circuits. We train AutoCkt over a sparse sub-sample of the design space which reduces convergence time during deployment towards reaching many new design specifications. AutoCkt has the following features:

\begin{description}

\item [$\bullet$] It intuitively understands the design space in the same manner as a circuit designer. Therefore, the framework is able to understand tradeoffs between different target specifications across the design space.
\item[$\bullet$] During run-time, it converges $\sim$40X faster than a traditional evolutionary algorithm. This allows the analog designer to quickly iterate through designs in an agile manner.
\item [$\bullet$] It reliably reaches many target specifications. In cases where AutoCkt fails, we show that these target specifications appear to be unreachable.
\item [$\bullet$] Using transfer learning, AutoCkt designs circuits while taking into account layout parasitics, 9.6X faster than the state-of-the-art \cite{Hakhamaneshi2019}.
\end{description}

We proceed to show our framework and results on three example circuits across different simulation environments including Spectre and the Berkeley Analog Generator, a tool that automatically simulates circuits  with layout parasitics.

\section{The Proposed Framework}
Figure \ref{system} shows the system level diagram for this algorithm; the two main blocks are the reinforcement learning agent and simulation environment, discussed further below.

\subsection{The Reinforcement Learning Agent}
Reinforcement Learning (RL) is a machine learning technique known to solve complex tasks in many systems. Specifically, it consists of an agent that iterates in an environment using a trial and error process that mimics learning in humans. It is a simulation-in-loop method, having the ability to verify outputs.

At each environment step, the RL agent, which contains a neural network, observes the state of the environment and takes an action based on what it knows. The environment then returns a new state that is used to calculate the reward for taking that particular action. The agent iterates through a trajectory of multiple environment steps, accumulating the rewards at each step until the goal is met or a predetermined maximum number of steps is reached. After running multiple trajectories the neural network is updated to maximize the expected accumulated reward via policy gradient.  

\begin{figure}[!t]
\centerline{\includegraphics{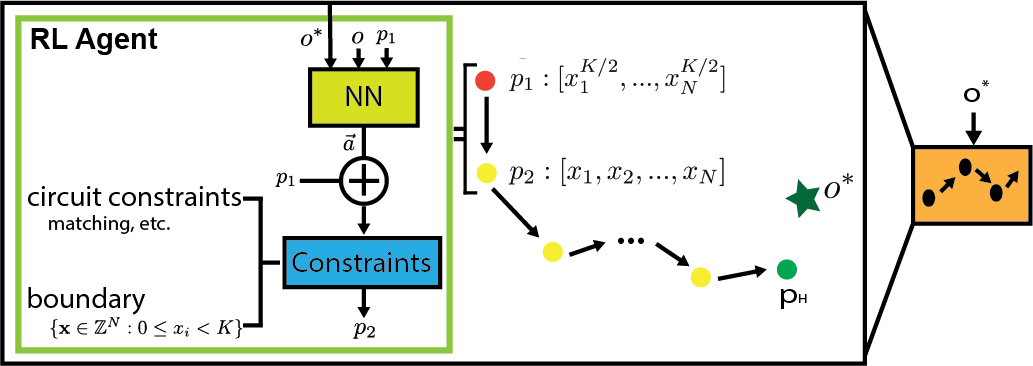}}
\caption{Trajectory generation showing how actions are taken by the reinforcement learning agent}
\label{trajectory}
\end{figure}
\begin{figure}[!t]
\centerline{\includegraphics{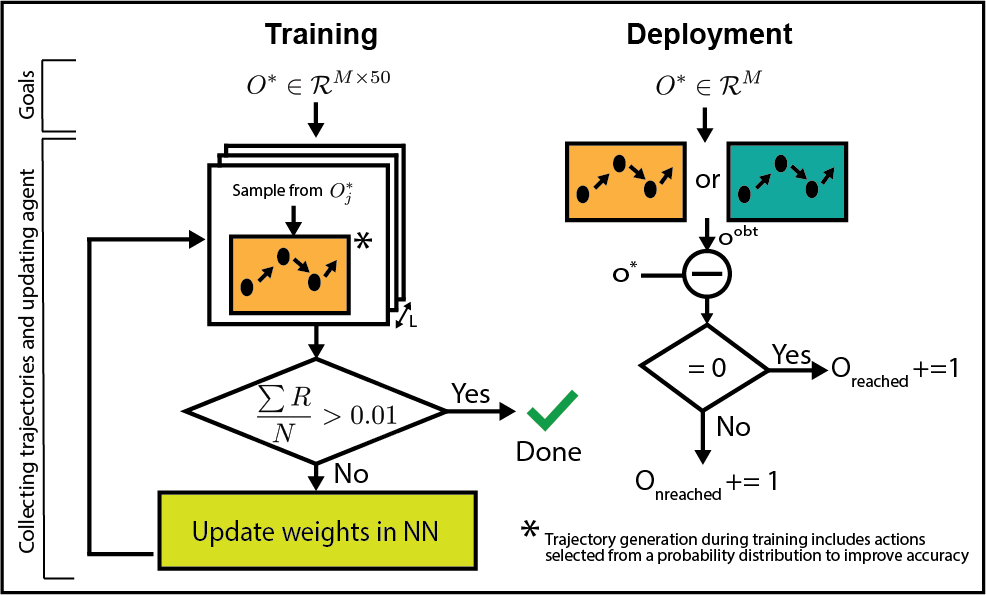}}
\caption{Total system level diagram of training and deployment process for AutoCkt}
\label{fc}
\end{figure}

In our application, there are $N$ parameters to tune for optimizing $M$ target design specifications. We can define our parameter space as $\mathbf{x} \in \mathbb{Z}^N$ and the design specifications space as $\mathbf{y} \in \mathbb{R}^M$, where $\mathbf{y}$ is normalized to a fixed range. The parameter space is originally a continuous space in $\mathbb{R}^N$ that is discretized to $K$ grids: $\{\mathbf{x} \in \mathbb{Z}^N : 0 \le x_i < K\}$.

\textbf{Trajectory Generation}
Figure \ref{trajectory} depicts how a trajectory is generated by AutoCkt. Upon reset, the parameters are initialized to the center point $\frac{K}{2}$. The neural network then uses the observed performance $o$ (created by simulating the circuit) and target specification $o^*$, as well as current parameters to decide whether to increment, decrement, or retain the same value for each circuit parameter. These actions are then constrained by any circuit specific rules or boundary limitations for the parameters. Note that these can be as specific or general as needed for different topologies, and AutoCkt is not reliant on having these circuit constraints exist. 

The agent has $H$ total simulation steps to reach $o^*$. If the objective is reached before $H$ steps, the trajectory ends. 

\textbf{Training and Deployment}
To train the RL agent, $50$ target specifications are randomly sampled: $$O^* = [o_i^* \in [o_i^{min}, o_i^{max}] \forall i \in [0,...,M]]\times 50$$ The number of target specifications needed to train was optimized through a hyperparameter sweep. $L$ trajectories are then generated, whose targets are chosen from $O^*$. The reward for each trajectory is obtained by accumulating the rewards for each action, formulated as a fairly typical dense reward:
$$R = \begin{cases} 
r, &\text{if } r < -0.01 \\
10 + r, &\text{if } r >= 0.01
\end{cases}
$$
where \begin{equation}\label{eq:rew}
\begin{aligned}
r = \sum_{i=1}^{M-T} min\{\frac{o_{pt\_i}-o^*_{pt\_i}}{o_{pt\_i}+o^*_{pt\_i}},0\} - \sum_{j=1}^{T} \epsilon\frac{o_{th\_j}-o^*_{th\_j}}{o_{th\_j}+o^*_{th\_j}}
\end{aligned}
\end{equation}

In Equation \ref{eq:rew}, $\mathbf{o_{pt}}$ represents hard constraint design specifications, and $\mathbf{o_{th}}$ represents design specifications that are being minimized. The reward increases as the RL agent's observed performance gets closer to the target specification. The training terminates once the mean reward has reached $0$, meaning all target specifications are consistently satisfied.

During deployment, the trained agent is used to generate trajectories with unique target specifications sampled from $O^*$. Note that the simulation environment can be different from the one used in training. The final $o$ obtained by the trajectory is then compared with $o^*$ and incremented in a respective counter. 

\begin{figure}[t!]
\centerline{\includegraphics{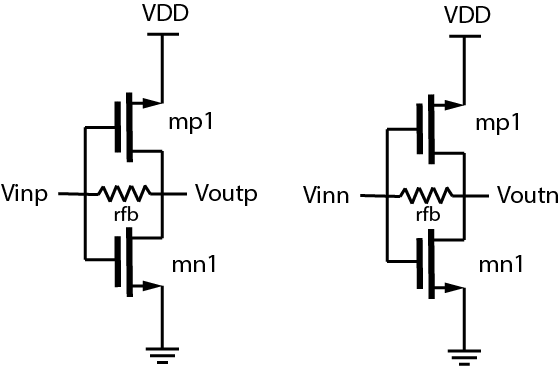}}
\caption{Simple transimpedance amplifier schematic}
\label{tia}
\end{figure}

\subsection{Simulation Environment}
AutoCkt is able to interface with different simulation environments. In this work, we demonstrate results using a simulator that works on predictive technology models and Spectre, which run on schematic level simulations, as well as the Berkeley Analog Generator (BAG), which runs simulations in Cadence with layout parasitics automatically.



\section{Experiments}
We demonstrate AutoCkt's capabilities with three different simulation environments as well as three circuit topologies. Each training session is conducted several times to ensure that AutoCkt is robust to variations in random seed. In our implementation, our neural network is a three layers with 50 neurons each, trained with Proximal Policy Optimization using OpenAI Gym and the Ray framework \cite{Liang2018} for running distributed reinforcement learning tasks. 

\subsection{Transimpedance amplifier}
\begin{figure}[t!]
\centerline{\includegraphics[scale=0.9]{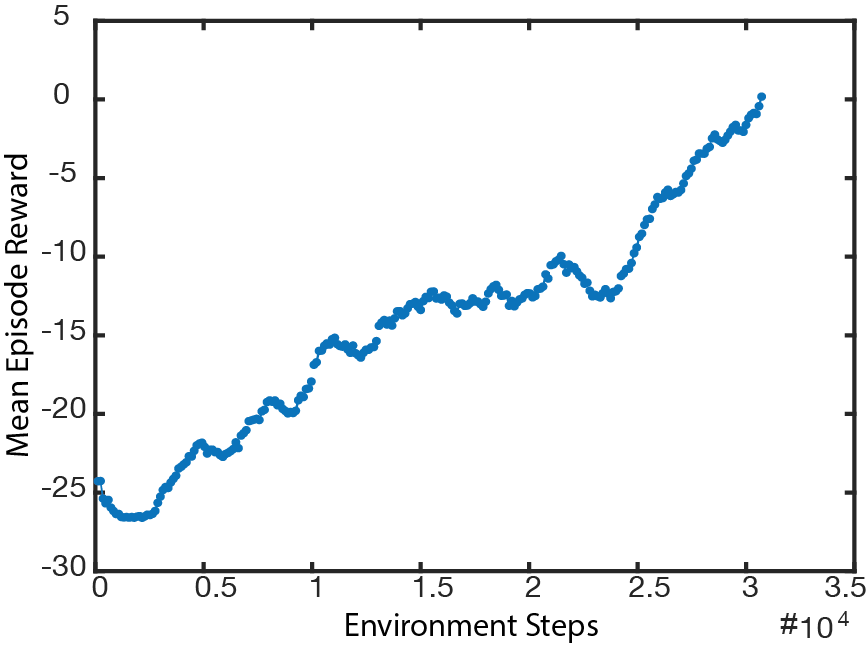}}
\caption{Mean episode reward for transimpedance amplifier}
\label{tia_rew}
\end{figure}

We first demonstrate AutoCkt's performance on a simple transimpedance amplifier (Figure \ref{tia}) in 45nm BSIM predictive technology. The action space for each transistor consists of two separate parameters (shown in array notation [start, end, increment]): width ($[2, 10, 2]*\mu m$) and multiplier ($[2, 32, 2]$). The feedback resistor action space consists of two parameters: number of resistors in series ($[2,20,2]$) and number of resistors in parallel ($[1,20,1]$). The fixed unit resistance is $5.6k\Omega$. The design specification space of interest is settling time ($[5,500]*ps$), cutoff frequency ($[5.0e^8, 7.0e^9]*Hz$) and input referred noise ($[100e^{-8}, 500e^{-6}]*V_{rms}$). Figure \ref{tia_rew} shows the mean episode reward over time increasing to greater than zero after training has completed, meaning that the agent has learned to reach the positive goal state across multiple target objectives.

\begin{table}[t!]
\caption{Sample Efficiency (SE) and Generalization Comparison Table: Transimpedance Amplifier}
\begin{center}
\begin{tabular}{|c|c|c|}
\hline
\textbf{Metric} & \textbf{TIA SE} & \textbf{Generalization TIA} \\
\hline
Genetic Alg. & 376 & N/A  \\
This Work & 15 & 487/500  \\
\hline
\end{tabular}
\label{tab4}
\end{center}
\end{table}

The trained agent was then deployed on 500 randomly chosen target specifications in the range specified above with results summarized in Table \ref{tab4}. The results show that AutoCkt has a 25.1$\times$ speedup compared to a vanilla genetic algorithm, as measured by sample efficiency which is the number of simulations it takes to converge to the target specification. Additionally, it is able to generalize to 97.4\% of the design space. Note the genetic algorithm efficiency was determined by the best result obtained when sweeping initial population sizes and several target specifications.

\subsection{Two stage operational amplifier}
\begin{figure}[t!]
\centerline{\includegraphics[scale=0.7]{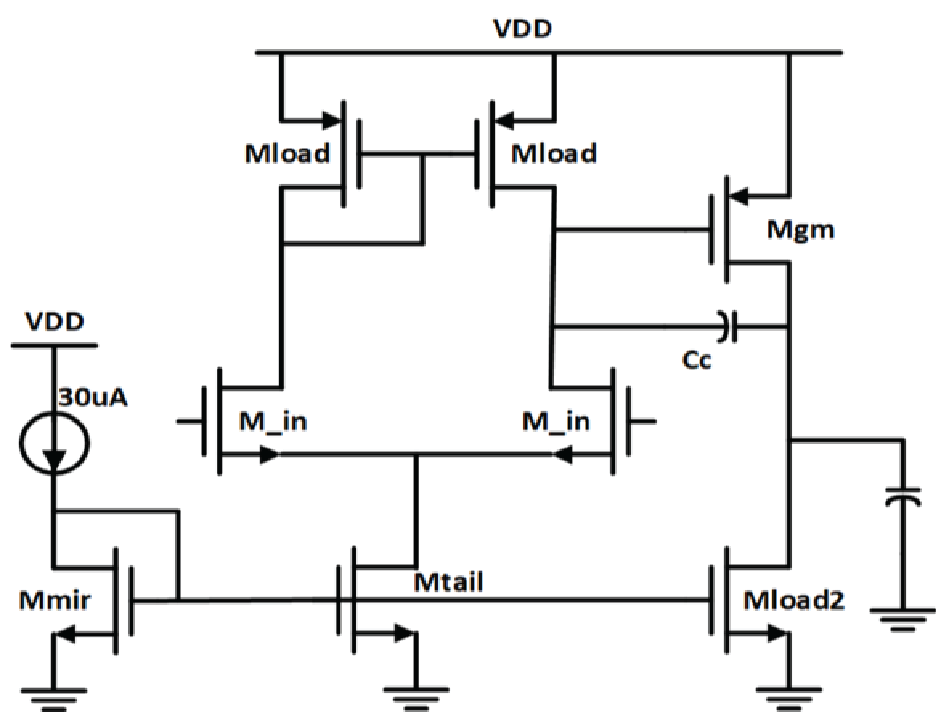}}
\caption{Two stage operational amplifier schematic}
\label{opamp}
\end{figure}

We move on to test AutoCkt on a more complex yet common circuit: a two stage operational amplifier (Figure \ref{opamp}) in 45nm BSIM predictive technology.

\begin{figure}[t!]
\centerline{\includegraphics[scale=0.8]{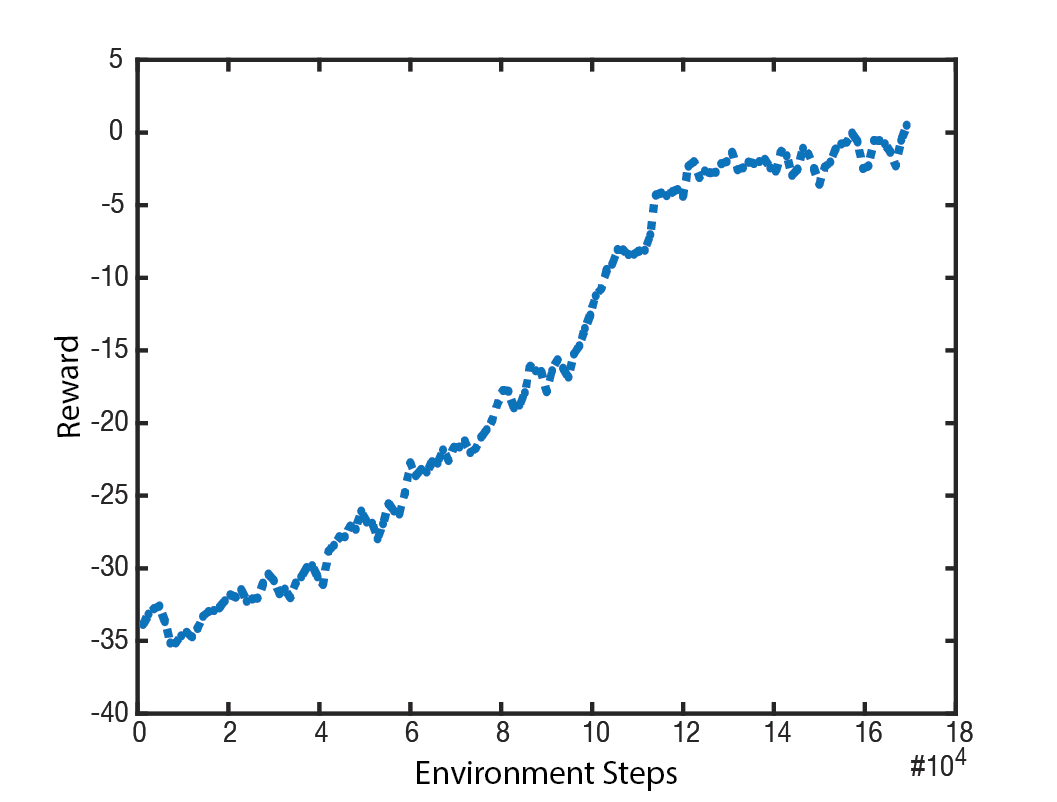}}
\caption{Mean reward over number of environment steps}
\label{lc}
\end{figure}

The action space for every transistor width in the schematic is $[1,100,1]*0.5\mu m$. The compensation capacitor ranges from $[0.1, 10.0, 0.1]*1pF$. The design specifications of interest are gain ($[200,400]*V/V$), unity gain bandwidth ($[1.0e^6,2.5e^7]*Hz$), phase margin ($[60.0]*^\circ$), and bias current (as a measure of power, $[0.1,10]*mA$). The total action space size is $10^{14}$ possible values, making random generation of parameters to meet the target design specification infeasible. The agent is allowed a trajectory length of $30$ simulation steps to converge. The mean reward over total environment steps is shown in Figure \ref{lc}.

\begin{figure}[t!]
\centerline{\includegraphics[scale=0.9]{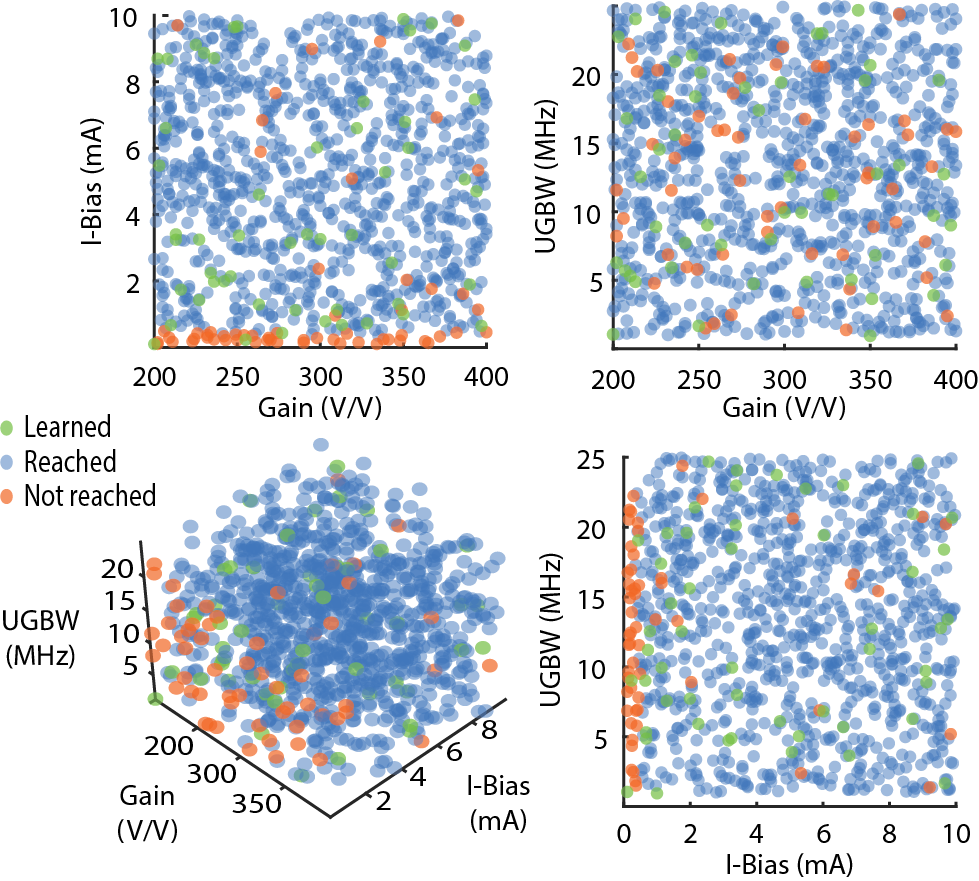}}
\caption{Distribution of learned, reached, and not reached target design specifications. Bottom left shows the 3D plot with three of the four design specifications. The rest of the plots show 2D plots for the differing combinations to demonstrate visually which points were not met.}
\label{distribution}
\end{figure}
We note that even though the agent took on the order $10^4$ steps to reach a mean reward of $0$, the amount of time to do just schematic simulation is 25 ms, making the overall training time tractable. We also utilize the capabilities of Ray \cite{Liang2018} to run multiple environments in parallel. Thus the wall clock time is just 1.3 hours on a 8 core CPU machine.

We run the trained agent on 1000 randomly generated target design specifications it has never seen before, in the range specified during training. The results are shown in a 3D plot (Figure \ref{distribution}, phase margin is excluded because it only has a lower bound requirement). The comparison table is shown in Table \ref{tab1}. Note that the comparison also includes a random RL agent taking steps in the environment, to illustrate design space complexity.  

The results demonstrate that AutoCkt is able to reach 963 of the 1000 target design specifications, generalizing by a factor of 20$\times$ compared to the specifications it saw during training. Of those points it does reach, the average number of simulation steps it takes is just 27, which is near 40$\times$ faster than a traditional genetic algorithm. In addition, the distribution of points in Figure \ref{distribution} show that the unreached design points fall along a vertical region where bias current is very low. We can then hypothesize that these points are indeed unreachable given the power requirement. Looking at the converged design specifications for these unreached points, we see that it attempts to meet the gain and bandwidth requirement while minimizing for power, similar to how a circuit designer approaches this problem. 

\begin{table}[t!]
\caption{Sample Efficiency (SE) and Generalization Comparison Table: Two Stage Op Amp}
\begin{center}
\begin{tabular}{|c|c|c|c|}
\hline
\textbf{Metric} & \textbf{Op Amp SE} & \textbf{TIA SE} & \textbf{Generalization Op Amp} \\
\hline
Genetic Alg. & 1063 & 376 & N/A  \\
Random RL Agent & N/A & N/A & 38/1000  \\
This Work & 27 & 15 & 963/1000  \\
\hline
\end{tabular}
\label{tab1}
\end{center}
\end{table}

\begin{figure}[b!]
\centerline{\includegraphics{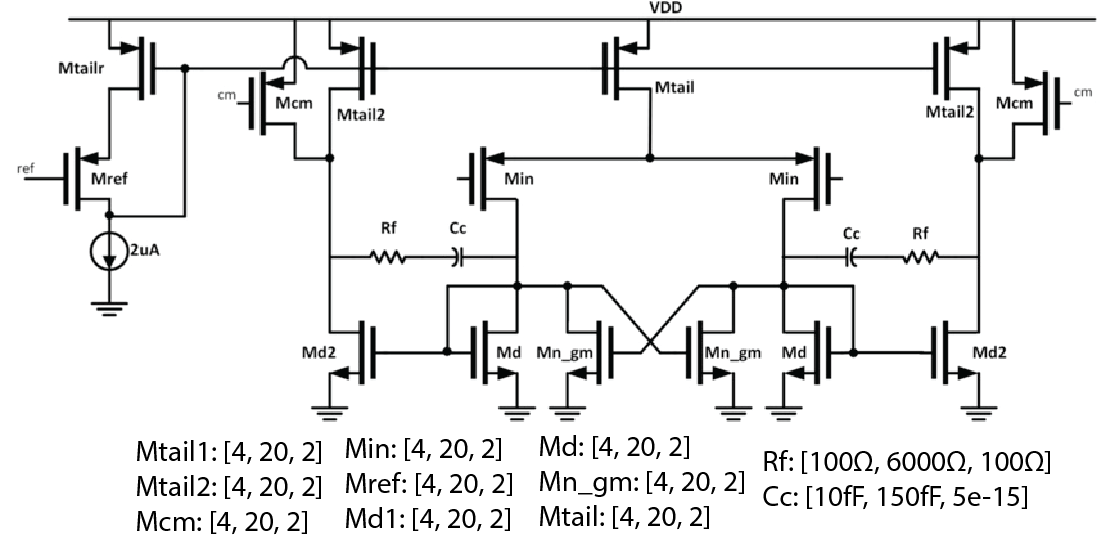}}
\caption{Schematic and action space for two stage op amp with negative $g_m$ load}
\label{negative_gm_schm}
\end{figure}

\begin{figure}[b!]
\centerline{\includegraphics[scale=0.9]{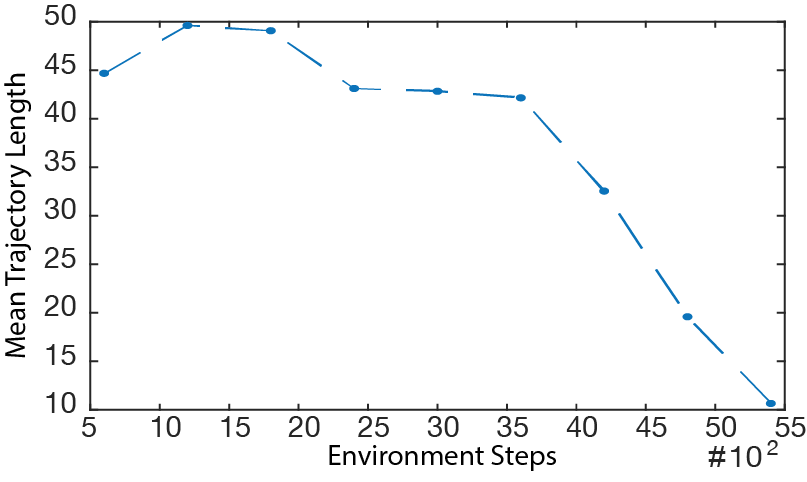}}
\caption{Trajectory length optimization for two stage op amp with negative $g_m$ load}
\label{eps_len}
\end{figure}

\subsection{Two stage OTA with negative $g_m$ load}
We demonstrate our algorithm functioning on an expert designed two stage operational amplifier with negative $g_m$ load in 16nm FinFet TSMC technology using Spectre. This circuit topology is shown in Figure \ref{negative_gm_schm}, and contains negative $g_m$ and diode-connected loads in the first stage, thereby having positive feedback, making the circuit more challenging to design and more sensitive to layout parasitics than a traditional amplifier.

\begin{figure}[t!]
\centerline{\includegraphics[scale=0.9]{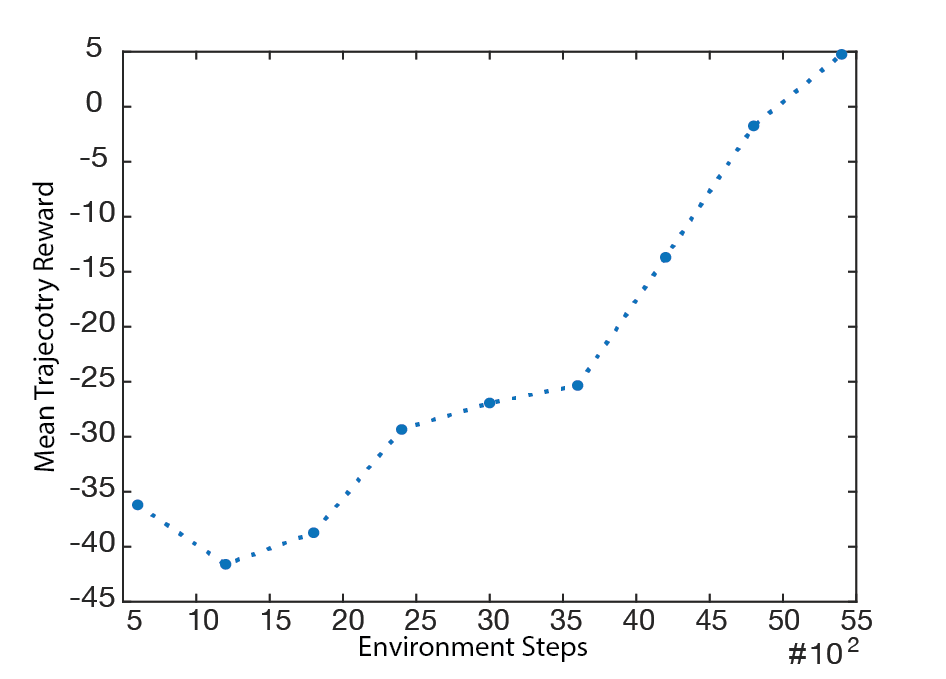}}
\caption{Mean episode reward over environment step for negative $g_m$ op amp}
\label{reward_neggm}
\end{figure}

\begin{figure}[b!]
\centerline{\includegraphics[scale=0.9]{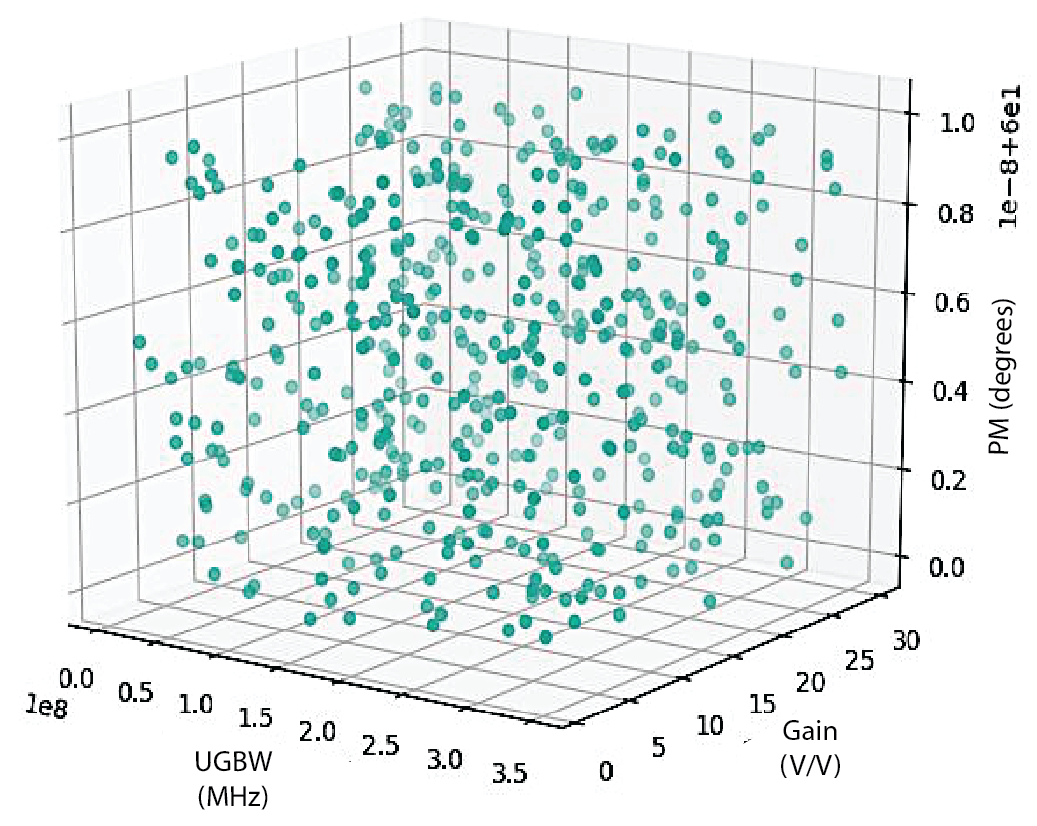}}
\caption{Distribution of reached target design specifications for the operational amplifier with negative $g_m$ load. Note that this example does not contain any unreached objectives.}
\label{negative_gm}
\end{figure}

The action space ranges are shown in the schematic, and the total order of complexity is $10^{11}$ different parameter combinations. The range for each design specification was chosen around an actual target design specification that the expert was trying to reach: gain ($[1,40]*V/V$), unity gain bandwidth ($[1.0e^6, 2.5e7]*Hz$), and phase margin ($[60,75]*^\circ$). The phase margin now includes a range; this is due to the transfer learning process to layout parasitics that will be presented later in this paper. The mean reward curve during training is shown in Figure \ref{reward_neggm}. Figure \ref{negative_gm} shows the results for 500 randomly generated target specifications after training the agent. Note that there are no unreached specifications.

The comparison table presented in Table \ref{tab2} shows very similar results compared to the prior two stage amplifier, with 40.6$\times$ faster convergence to a target specification compared to a traditional genetic algorithm, taking on average just 10 simulations to converge to a solution (see Figure \ref{eps_len}).

\begin{table}[t!]
\caption{Sample Efficiency (SE) and Generalization Comparison Table: Two Stage Op Amp with Negative $g_m$ Load}
\begin{center}
\begin{tabular}{|c|c|c|c|}
\hline
\textbf{Metric} & \textbf{Op Amp SE} & \textbf{Generalization Op Amp} \\
\hline
Genetic Alg. & 406 & N/A  \\
Random RL Agent & N/A & 4/500  \\
This Work & 10 & 500/500  \\
\hline
\end{tabular}
\label{tab2}
\end{center}
\end{table}

\subsection{Two stage operational amplifier with negative $g_m$ load and layout parasitics}

Most prior analog sizing tools lack the capability of sample efficiently considering post-layout extracted (PEX) simulations, due to the lack of automatic generation of layout. Leveraging the Berkeley Analog Generator (BAG) \cite{Chang2018}, we can encapsulate an expert designer's layout methodology to generate layouts across a comprehensive set of input parameters. In our framework, we also consider different PVT variations, taking the worst performing metric as the specification. The entire simulation process, however, takes significantly more time: the schematic simulation for the two stage op amp in Figure \ref{negative_gm_schm} takes just $2.4$ seconds, whereas including layout parasitics in BAG takes, on average, $91$ seconds to complete. The almost 38$\times$ factor in simulation time implies prior work cannot scale to more complex topologies due to inaccuracy or sample inefficiency. 

\begin{figure}[t!]
\centerline{\includegraphics[scale=0.9]{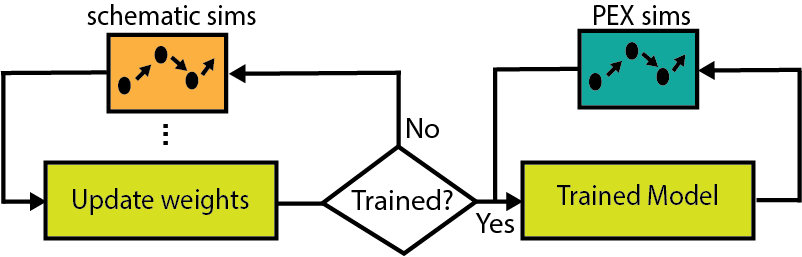}}
\caption{Diagram illustrating the transfer learning process in order to run PEX simulations}
\label{transfer}
\end{figure}

We demonstrate the usage of transfer learning to show that an RL agent trained by running inexpensive schematic simulations is able to transfer it's knowledge to a different environment. This new environment, which then runs PEX simulations, is then used to deploy the agent. Figure \ref{transfer} shows this idea. Note that no training is done once the environment has changed to post-layout extraction.

To demonstrate transfer learning, the agent trained on the two stage op amp with negative $g_m$ load in Spectre is then run on the TSMC 16nm FF operational amplifier generator in BAG. The target design specifications are randomly chosen within the same range as the schematic-trained agent with the exception of phase margin, where we only enforce a minimum requirement of $60^\circ$. In our tests, we found that training on a range of phase margins, as opposed to a single lower bound of $60^\circ$, resulted in a better transfer performance. This is likely due to the agent benefiting from more exploration of the design space.

\begin{figure}[t!]
\centerline{\includegraphics[scale=0.8]{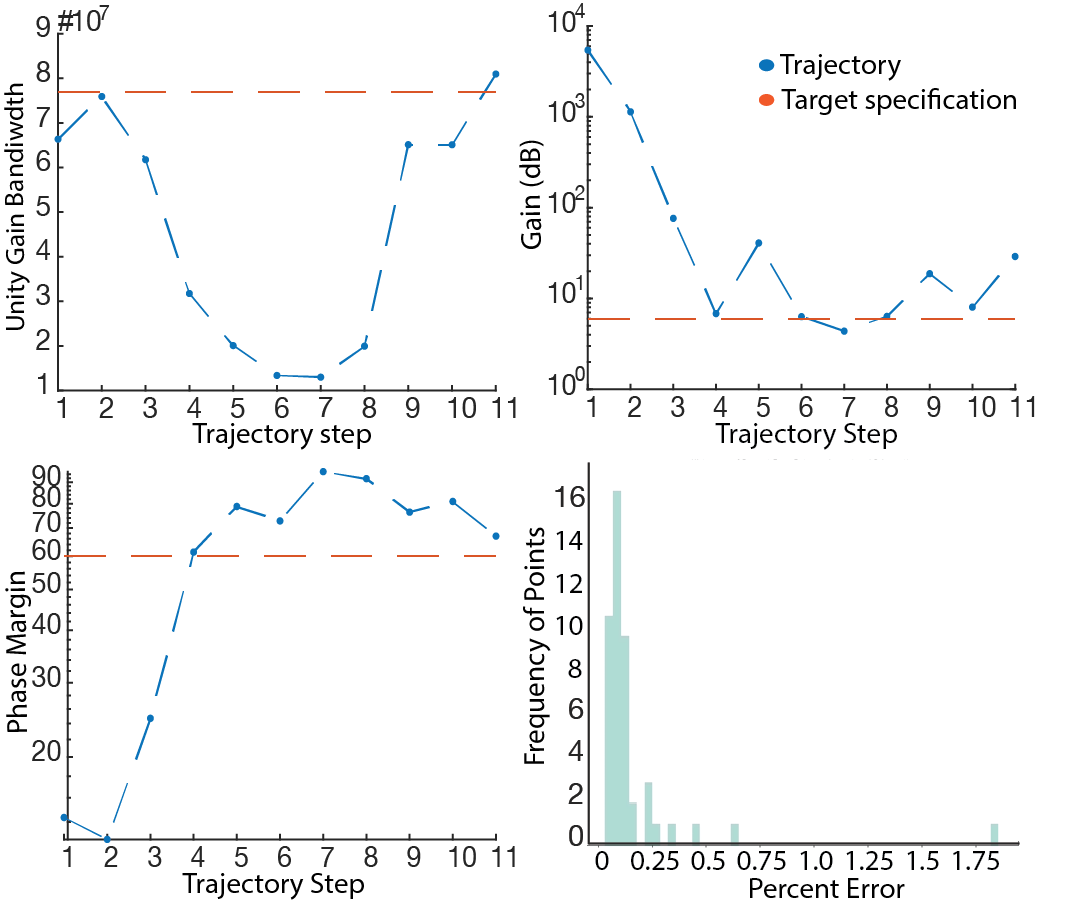}}
\caption{Top left, top right, and bottom left figures show a sample trajectory for the transferred agent attempting to reach one target design specification. Bottom right shows a histogram plotting difference between schematic and layout simulation.}
\label{traj}
\end{figure}

A sample trajectory for a single target design specification is shown in Figure \ref{traj}. These trajectories illustrate that in 11 time steps, the agent is able to converge to a target design objective that does indeed meet specification. 

In general, compared to it's schematic counterpart, the transferred agent takes longer to converge to a design that meets the target specification (shown in Table \ref{tab3}) due to the addition of layout parasitics. Figure \ref{traj} shows a histogram of 50 design points that calculate the average percent difference across each design specification between PEX and schematic simulation. We posit that the agent learns the intuitive tradeoffs between parameters and design specifications as well as the best actions to take to move towards a goal, and that these relationships hold when considering layout parasitics despite potentially large amounts of difference between the schematic and PEX simulations. 

Table \ref{tab3} shows that running a vanilla genetic algorithm is too sample inefficient. We also compare AutoCkt to the combined machine learning and genetic algorithm \cite{Hakhamaneshi2019} and show that the sample efficiency of AutoCkt is 9.56$\times$ greater than the prior state-of-the-art. Running on a single core CPU, our algorithm takes just 1.7 hours to complete. We run the algorithm on 40 randomly generated target design specifications, and AutoCkt is able to to obtain 40 LVS passed designs in under three days, with no parallelization.     

\begin{table}[t!]
\caption{Sample Efficiency (SE) and Generalization Comparison Table: Two Stage Op Amp with Negative $g_m$ Load and Layout Parasitics}
\begin{center}
\begin{tabular}{|c|c|c|c|c|}
\hline
\textbf{Metric} & \textbf{Sim Steps} & \textbf{Generalization} \\
\hline
Genetic Alg. & N/A & N/A  \\
Genetic Alg.+ML \cite{Hakhamaneshi2019} & 220 & N/A  \\
AutoCkt Schematic Only  & 10 & 500/500\\
AutoCkt PEX & 23& 40/40  \\
\hline
\end{tabular}
\label{tab3}
\end{center}
\end{table}

\section{Conclusion}
In this paper, we present a machine learning framework that designs analog circuits. Compared to prior optimization approaches, AutoCkt is on average 40$\times$ more sample efficient than a genetic algorithm. We demonstrate the robustness of our framework on three circuit topologies in different simulation environments. By leveraging transfer learning, AutoCkt considers layout parasitics, and is 9.6$\times$ more sample efficient than the state-of-the-art. We show that using only a 1 core CPU, our algorithm is able to design 40 LVS passing designs for two stage OTA with negative $g_m$ load in under 3 days.

\section{Acknowledgments}
This work is supported by DARPA CRAFT (HR0011-16-C-0052), ADEPT, and BWRC member companies.

\bibliographystyle{ieeetr}
\bibliography{bib}

\blfootnote{Open-sourced code can be found at: https://github.com/ksettaluri6/AutoCkt}
\end{document}